\documentclass[prb,aps,twocolumn,superscriptaddress,showpacs]{revtex4-1}

\usepackage[colorlinks=true,linkcolor=black, citecolor=blue, urlcolor=blue, 
    unicode=true]{hyperref}
\usepackage{color}
\usepackage{graphicx}
\usepackage{amsmath}
\usepackage{blindtext}
\usepackage{gensymb}
\usepackage{bm}
\usepackage[version=4]{mhchem}

\begin{document}


\title{Magnetic surface reconstruction in the van-der-Waals antiferromagnet Fe$_{1+x}$Te}

\author{C. Trainer}
\affiliation{School of Physics and Astronomy, University of St. Andrews, St. Andrews KY16 9SS, UK}
\author{M. Songvilay}
\affiliation{School of Physics and Astronomy, University of Edinburgh, Edinburgh EH9 3JZ, UK}
\author{N. Qureshi}
\affiliation{Institut Laue-Langevin, 71 avenue des Martyrs, CS20156, 38042 Grenoble Cedex 9, France}
\author{A. Stunault}
\affiliation{Institut Laue-Langevin, 71 avenue des Martyrs, CS20156, 38042 Grenoble Cedex 9, France}
\author{C. M. Yim}
\affiliation{School of Physics and Astronomy, University of St. Andrews, St. Andrews KY16 9SS, UK}
\author{E. E. Rodriguez}
\affiliation{Department of Chemistry and Biochemistry, University of Maryland, College Park,Maryland 20742, USA}
\author{C. Heil}
\affiliation{Institute of Theoretical and Computational Physics, Graz University of Technology, NAWI Graz, 8010 Graz, Austria}
\author{V. Tsurkan}
\affiliation{Center for Electronic Correlations and Magnetism, Experimental Physics V, University of Augsburg, D-86159 Augsburg, Germany}
\affiliation{Institute of Applied Physics, Academy of Sciences of Moldova, MD 2028 Chisinau, Republic of Moldova}
\author{M. A. Green}
\affiliation{School of Physical Sciences, University of Kent, Canterbury, CT2 7NH, UK}
\author{A. Loidl}
\affiliation{Center for Electronic Correlations and Magnetism, Experimental Physics V, University of Augsburg, D-86159 Augsburg, Germany}
\author{P. Wahl}
\affiliation{School of Physics and Astronomy, University of St. Andrews, St. Andrews KY16 9SS, UK}
\author{C. Stock}
\affiliation{School of Physics and Astronomy, University of Edinburgh, Edinburgh EH9 3JZ, UK}

\date{\today}

\begin{abstract}

Fe$_{1+x}$Te is a two dimensional van der Waals antiferromagnet that becomes superconducting on anion substitution on the Te site.  The properties of the parent phase of Fe$_{1+x}$Te are sensitive to the amount of interstitial iron situated between the iron-tellurium layers.  Fe$_{1+x}$Te displays collinear magnetic order coexisting with low temperature metallic resistivity for small concentrations of interstitial iron $x$ and helical magnetic order for large values of $x$.  While this phase diagram has been established through scattering [see for example E. E. Rodriguez $\textit{et al.}$ Phys. Rev. B ${\bf{84}}$, 064403 (2011) and S. R\"o\ss{}ler $\textit{et al.}$ Phys. Rev. B ${\bf{84}}$, 174506 (2011)], recent scanning tunnelling microscopy measurements [C. Trainer $\textit{et al.}$ Sci. Adv. ${\bf{5}}$, eaav3478 (2019)] have observed a different magnetic structure for small interstitial iron concentrations $x$ with a significant canting of the magnetic moments along the crystallographic $c$ axis of $\theta$=28 $\pm$ 3$^{\circ}$.  In this paper, we revisit the magnetic structure of Fe$_{1.09}$Te using spherical neutron polarimetry and scanning tunnelling microscopy to search for this canting in the bulk phase and compare surface and bulk magnetism.  The results show that the bulk magnetic structure of Fe$_{1.09}$Te is consistent with collinear in-plane order ($\theta=0$ with an error of $\sim$ 5$^{\circ}$). Comparison with scanning tunnelling microscopy on a series of Fe$_{1+x}$Te samples reveals that the surface exhibits a magnetic surface reconstruction with a canting angle of the spins of $\theta=29.8^{\circ}$.  We suggest that this is a consequence of structural relaxation of the surface layer resulting in an out-of-plane magnetocrystalline anisotropy.  The magnetism in Fe$_{1+x}$Te displays different properties at the surface when the symmetry constraints of the bulk are removed. 
\end{abstract}


\maketitle

\section{Introduction}

Van der Waals forces differ from ionic and covalent bonding in terms of the strength and also the range of forces.~\cite{Mayor16:29}  Two dimensional materials that are based on sheets weakly held together by van der Waals forces have recently been studied in the context of graphene~\cite{Neto09:81}, and also for the investigation of two dimensional ferromagnetism~\cite{Miller17:70}. Two dimensional magnetic van der Waals crystals have also been of interest in the context of iron based superconductivity~\cite{Ishida09:78,Kamihara08:130,Johnson1010:59,Paglione10:6,Dai15:87,Inosov15:59,Stock10:6,Dai12:8,Lumsden10:22} with arguably the structurally simplest such superconductor being the monolayer compound \cite{Hsu08:38} Fe$_{1+x}$Te$_{1-y}$(Se,S)$_{y}$~\cite{Wen11:74} consisting of weakly bonded iron chalcogenide sheets.  Under anion substitution, an optimal superconducting transition temperature of $\sim 14\mathrm{K}$ has been reported in Fe$_{1+x}$Te$_{0.5}$Se$_{0.5}$~\cite{Sales09:79} and $\sim 10\mathrm{K}$ in FeTe$_{1-x}$S$_{x}$~\cite{Mizuguchi09:94}.  In this paper we investigate the difference between the bulk and surface magnetic structures in the non-superconducting parent compound Fe$_{1+x}$Te through the use of spherical neutron polarimetry and  spin polarized scanning tunnelling microscopy in vector magnetic fields.

The single layered chalcogenide Fe$_{1+x}$Te$_{1-y}$Q$_{y}$ (where $Q$=Se or S) has been important in the study of iron based superconductors owing to its relatively simple single layer structure and because it is highly electronically localized~\cite{Yin111:10,Rossler10:82,Si08:101,Si09:5} in comparison to other iron based systems.  This is a property that is also reflected in the oxyselenides.~\cite{McCabe14:89,Zhao13:87,Zhu10:104,Freelon15:92}  The electronic and magnetic properties of Fe$_{1+x}$Te$_{1-y}$Q$_{y}$ can be tuned via two variables - the parameter $x$ determines the amount of interstitial iron located between the weakly bonded FeTe layers and disordered throughout the crystal, and $y$ is the amount of anion substitution and provides a chemical route towards superconductivity.  It should be noted that while the interstitial iron is disordered introducing magnetic clusters~\cite{Thampy12:108}, the electronic properties have been found to be homogeneous,~\cite{He11:83,Xu18:97} and recently discussed in the context of device fabrication~\cite{Zalic19:100}.  The sensitivity of the properties to stoichiometry is also reflected in Fe$_{1+\delta}$Se.~\cite{McQueen09:79}  There have been several studies that have shown $x$ and $y$ to be correlated and hence both influence the superconductivity.~\cite{Rodriguez11:2,Sun16:6,Babkevich10:22,Bendele10:82}  In particular, the tetrahedral bond angles~\cite{Huang10:104,Xu16:93,Xu10:82,Wen12:86} are altered with interstitial iron concentration $x$ along with tuning the material across several magnetic and structural phase transitions. It is this interplay between the structural, magnetic, electronic and superconducting properties which makes this material exciting, with the expectation that understanding the relation between these phases leads to an understanding of superconductivity.

Fe$_{1+x}$Te has been found to display two spatially long-range correlated magnetic phases as a function of iron concentration $x$ separated by a region of spatially short-ranged magnetic order~\cite{Bao09:102,Koz13:88,Wen09:80}.   For small interstitial iron concentrations $x\leq 0.12$, the magnetic structure is collinear~\cite{Li09:79,Zal12:85} below a temperature where a structural transition occurs from a tetragonal ($P4/nmm$) to monoclinic ($P2_{1}/m$) unit cell.  Electronically, this also marks a transition from a resistivity which is ``semi/poor"-metallic to being metallic in character at low temperatures.~\cite{Chen09:79}  Despite the metallic character, the low energy spin fluctuations are consistent with localized transverse spin-waves~\cite{Stock17:95,Song18:97}.  The second magnetic phase at concentrations $x>0.12$ is helical in nature combined with a ``semi/poor"-metallic behavior at all temperatures.~\cite{Rossler11:84}  The two disparate magnetic phases induced with the variable $x$ are separated by a collinear spin density wave~\cite{Materne15:115,Parshall12:85} located near a Lifshitz point~\cite{Rodriguez13:88}.  As well as the tuning of the magnetic and crystallographic structures with interstitial iron concentration, the magnetic excitations also display a large dependence on $x$~\cite{Stock11:84,Stock12:85}, even at high energy transfers~\cite{Lumsden10:6,Stock90:14,Zal11:107,Xu16:93,Lipscombe11:106}.

We will focus on the low interstitial iron concentrations in this study.  The magnetic structure for low interstitial iron concentrations is termed a ``double-stripe" structure and has been investigated extensively with both unpolarized and uniaxial polarized neutron scattering.  This collinear magnetic phase has magnetic moments aligned along the crystallographic $b$ axis and magnetic Bragg peaks in the neutron cross section at $\vec{Q}$=$({1\over 2}, 0, {1\over 2})$ or denoted as $(\pi, 0)$.  Being metallic at low temperatures and easily cleavable offers the opportunity to apply surface techniques for investigating the electronic and magnetic properties.  

In particular, spin-polarized scanning tunnelling microscopy has been used to manipulate the excess iron concentration at the surface layer.~\cite{Enayat14:345,Singh15:91}  This work has demonstrated that the surface magnetic structure faithfully follows the bulk magnetic phase diagram as a function of interstitial iron in terms of both the crystallographic and magnetic structures giving consistent results with neutron scattering on the interstitial iron concentration where the magnetic and crystallographic structures change. However, an important difference was observed for the magnetic structure in the collinear ``double-stripe" phase for small values of $x$.  Tunneling measurements show a periodicity consistent with the stripe phase reported based on neutron scattering~\cite{Sugimoto13:45}, however recent measurements in vector magnetic fields have observed a significant out of plane canting, along the crystallographic $c$-axis, of the magnetic moment of $\theta \sim 28^\circ$ where neutron scattering reports the moments to be entirely in the $ab$ plane ($\theta=0$).~\cite{Hanke2017,Trainer19:5}  Interestingly, such a magnetic structure is consistent with early studies on Fe$_{1.12}$Te,\cite{Fruchart75:10} however given more recent work it is possible that this concentration was at the boundary between collinear and helical magnetism possibly complicating the interpretation.

\begin{figure}[t]
\includegraphics[width=85mm]{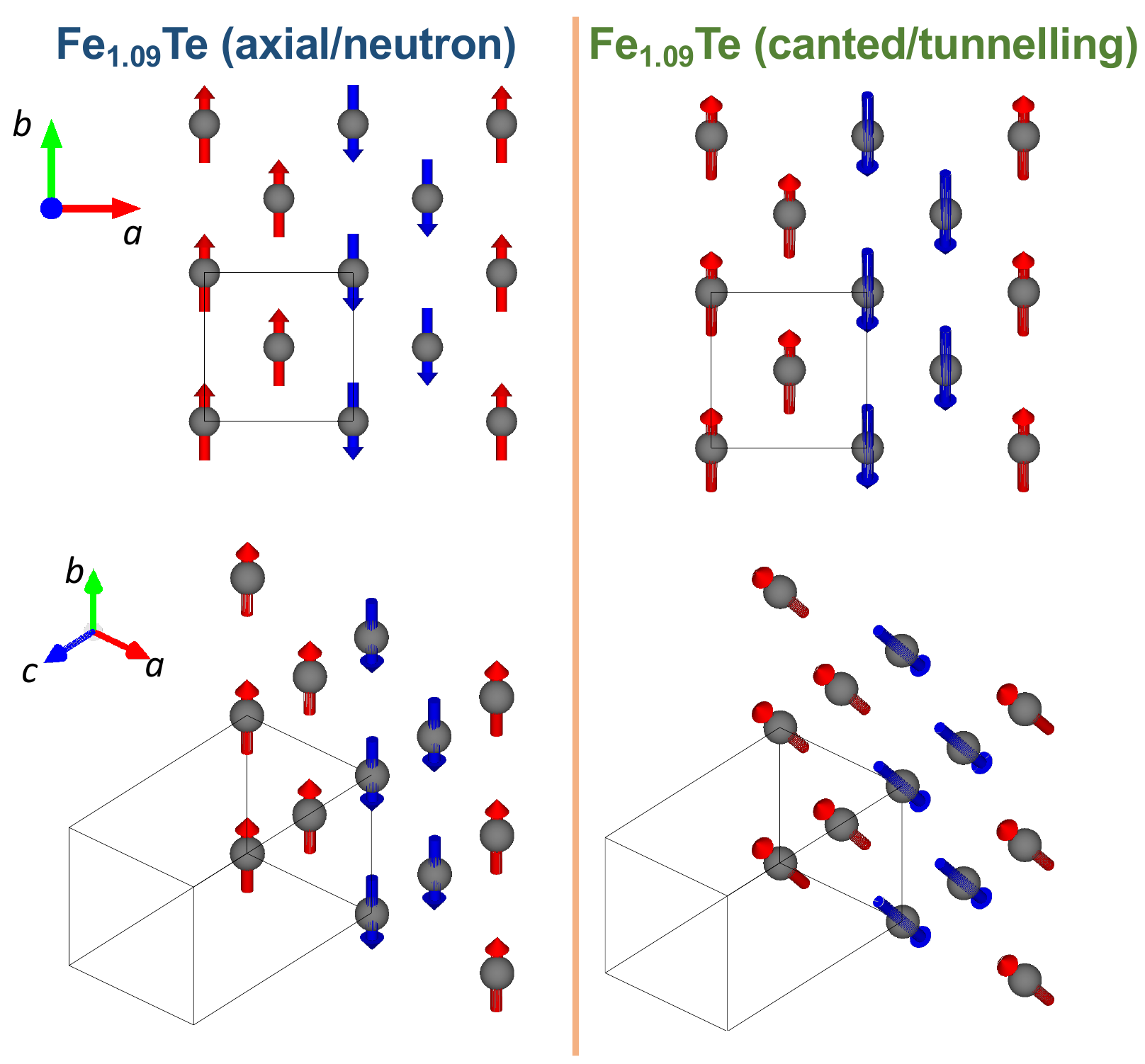}
\caption{Comparison of magnetic order from neutron scattering and STM. The axial magnetic structure with the moments aligned along the crystallographic $b$ axis ($\theta$=0) contrasts with the canted structure measured with tunnelling measurements where the moments are canted out of the plane by $\theta$=28 $\pm$ 3$^{\circ}$. } \label{fig:structure}
\end{figure}

Based on the discrepancy between current neutron diffraction results, scanning tunneling microscopy, and older diffraction work, we revisit this problem applying spherical neutron polarimetry to determine the out of plane angle due to any canting of the spins in bulk single crystals of Fe$_{1+x}$Te. This study focuses on the iron deficient portion of the phase diagram which exhibits collinear order because this is where the differences between neutron scattering and STM are most prominent. 

In this paper, we compare spin polarized scanning tunnelling microscopy measurements of the magnetism on the surface with a study of the bulk magnetic structure.  The two different magnetic structures that will be compared in this paper are illustrated in Fig. \ref{fig:structure}.  This paper is divided into five sections including this introduction.  We first present the results from spin polarized scanning tunnelling microscopy of the canting angle in the surface layer. We then investigate the canting angle in the bulk from spherical neutron polarimetry and analyze the results in terms of a possible canting in the bulk. We finally compare these results and discuss the differences and possible origins, including dipolar and anisotropy terms in the magnetic Hamiltonian.  Through this comparison we find that the surface layer of the two dimensional van der Waals Fe$_{1+x}$Te magnet exhibits a magnetic surface reconstruction.

\section{Spin Polarized STM measurement}

\begin{figure*}[t]
\includegraphics[width=170mm]{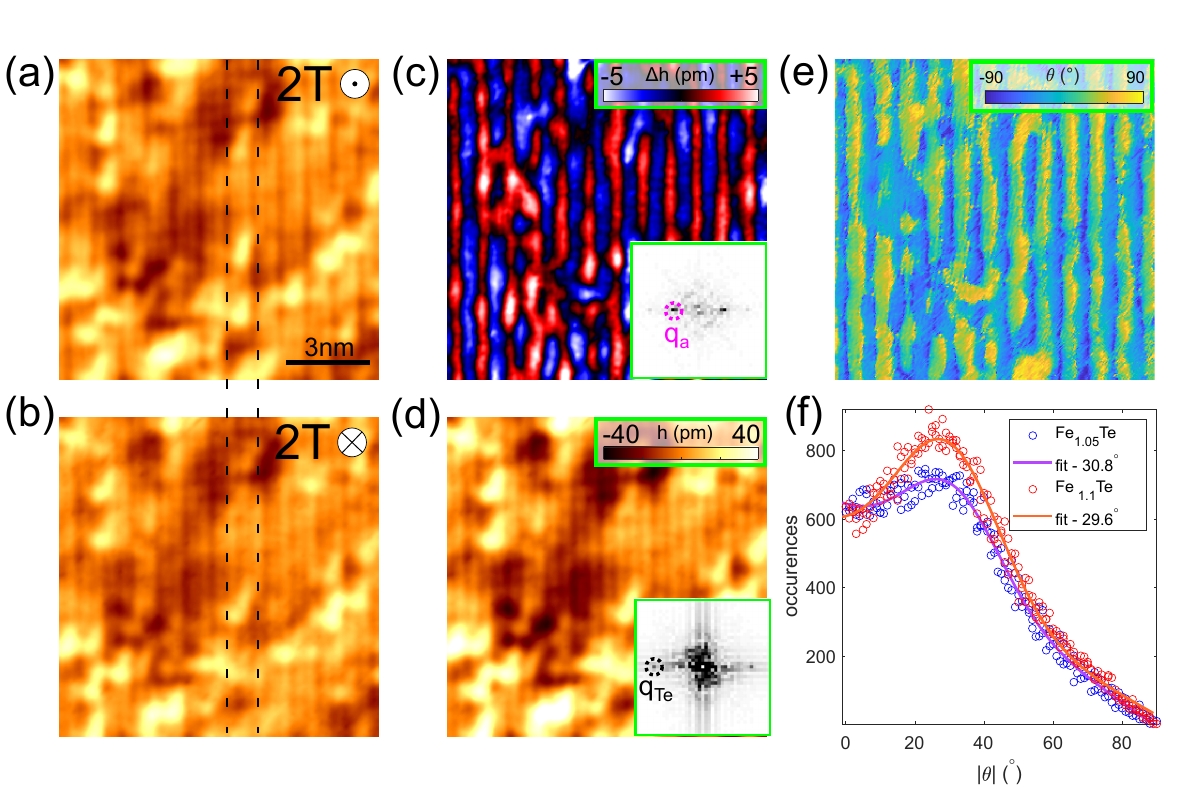}
\caption{(a) - Spin polarized STM image of the surface of an $\mathrm{Fe}_{1.1}\mathrm{Te}$ sample with a bias voltage of $100\mathrm{mV}$ and a set point current of $50\mathrm{pA}$. Recorded with an applied magnetic field of $\mathrm{2T}$ applied out of the plane of the image along the sample $c$ axis. (b) As (a) but with the direction of the applied magnetic field reversed. (c) Half the difference of the images shown in (a) and (b), directly proportional to the sample magnetization along the $c$ axis. Inset - The Fourier transform of (c) showing the magnetic ordering vector ($q_a$). (d) The average of the images (a) and (b), directly proportional to the non spin polarized component of the tunneling current. Inset - the Fourier transform of (d) showing the atomic peaks due to the $\mathrm{Te}$ lattice. (e) The out of plane canting angle ($\theta$) measured from images recorded with three orthogonal directions of applied field. (f) Histograms of the absolute value of the measured out of plane canting angle for data sets of three dimensional spin polarization data recorded on different samples of $\mathrm{FeTe}$. The red data points corresponds to the data shown in (e). Solid lines represent fits of a Gaussian function plus a linear background.  } \label{fig:spstm}
\end{figure*}

We first discuss spin polarized STM measurements of Fe$_{1+x}$Te probing the magnetic structure at the surface.  

\subsection{Experimental Details}
Spin polarized STM measurements were conducted on samples of Fe$_{1+x}$Te with excess iron concentrations $x$ ranging from $5\%$ to $11.5\%$. The measurements were performed using a home-built cryogenic STM operating at a base temperature of $2\mathrm{K}$ and mounted in a Vector magnet that is capable of applying a field of up to $5\mathrm{T}$ in any direction relative to the sample~\cite{Trainer2017}. Atomically clean surfaces for STM measurement were prepared by cleaving the samples in-situ at a temperature of $\sim 20 \mathrm{K}$ \cite{white_stiff_2011}. Magnetic tips were created by collecting excess Fe atoms from the sample surface \cite{Enayat14:345,Trainer19:5,Singh15:91}. In this way, the tunneling current between the STM tip and sample becomes sensitive to the relative angle between the magnetization of the tip and the sample. The tunneling current ($I_\mathrm{SP}$) due to the spin polarization of the tip ($P_{\mathrm{tip}}$) and the sample ($P_{\mathrm{sample}}$) can be expressed as\cite{Wiesendanger2009} 

\begin{equation}
  I_{\mathrm{SP}}=I_0\left[1+P_{\mathrm{tip}}P_{\mathrm{sample}}\cos(\phi)\right],
  \label{sp-stmeqn}
\end{equation}
 
\noindent where $\phi$ is the angle between the tip and sample magnetizations.  Figure~\ref{fig:spstm}(a) shows a typical spin polarized STM topographic image of the iron telluride surface (Fe$_{1.1}$Te). The excess iron atoms are seen in the STM images as bright protrusions on the surface. The bi-collinear antiferromagnetic order is imaged as a stripe like modulation running parallel to the sample $b$ axis with a wavevector $(q_a)$ along the sample $a$ axis. The imaged wavelength and direction of this ordering is in excellent agreement with that determined from neutron scattering\cite{Bao09:102,Rodriguez11:84}. For ferromagnetic tips, the magnetization of the tip is found to follow the direction of an applied magnetic field \cite{Enayat14:345,Trainer19:5,Hanke2017} therefore imaging the surface with the tip polarized by a field applied $180\degree$ to the original field orientation results in a $\pi$ phase reversal of the imaged magnetic order. This reversal of the imaged magnetic order can be seen by comparing images recorded with opposite applied field orientations shown in Figs.~\ref{fig:spstm}(a) and (b).

\subsection{STM Results}
It is possible, through Eqn.~\ref{sp-stmeqn}, to directly measure the sample's surface spin polarization from the spin polarized STM images. This is done by taking the difference of the images recorded with oppositely polarized tips which is proportional to $2P_{\mathrm{tip}}P_{\mathrm{sample}}\mathrm{cos}(\phi)$. Fig. \ref{fig:spstm}(c) shows such a difference image showing the component of the sample's magnetic order that is parallel to the crystal $c$ axis. The sum of the two images recorded with oppositely polarized tips resembles the topography that would be recorded if the sample was imaged with a non-spin polarized tip (Fig.~\ref{fig:spstm}(d)). 
 
By recording spin-polarized STM images with magnetic field applied in three orthogonal spatial directions it is possible to determine the precise orientation of the sample's spin structure at the surface \cite{Zhang2016, Trainer19:5}. The out of plane canting angle ($\theta$) of the surface spins resulting from this measurement is shown in Fig. \ref{fig:spstm}(e). Clear canting of the spins away from the $ab$ plane can be observed. By plotting the absolute value of this angle as a histogram, shown in Fig.~\ref{fig:spstm}(f), a clear peak at $\sim30\degree$ can be observed. This measurement has been repeated for a sample of Fe$_{1.05}$Te, the results of which are also shown in Fig.~\ref{fig:spstm}(f).  We determine the out of plane canting of the surface spins by fitting a Gaussian distribution plus a linear background to the data. By combining the fits to both data sets we obtain an average out of plane canting angle of $30.3\pm2.0^\circ$. This substantial canting of the surface spins is seen across multiple samples and has been observed in previous STM studies on this compound\cite{Hanke2017,Trainer19:5}.  
  
 \begin{table}[t]
\begin{tabular}{c|c|c|c|c}
  sample \# & $x (\%)$& $\mathrm{I}(q_a) ||b$ & $\mathrm{I}(q_a) ||c$ & $\theta (\degree)$ \\
\hline
1 &5 &0.1221      & 0.0550      & 24.2527         \\
2 &10 &0.2246      & 0.1570      & 34.9522         \\
2 &10 &0.1542      & 0.1721      & 48.1350         \\

2 &10 &0.2719      & 0.0186      & 3.9146          \\
2 &10 &0.2243      & 0.1761      & 38.1379         \\
3 &11.5 &0.2248      & 0.1275      & 29.5626         \\

\end{tabular}
\caption{Results from $6$ different independent SP-STM measurements on different samples of Fe$_{1+x}$Te with different magnetic tips. The table shows the intensity of the imaged magnetic order when the tip spin is parallel to the crystal $b$ axis and when the tip spin is parallel  to the crystal $c$ axis and the resulting canting angle of the surface spins.}
\label{table3}
\end{table}
 
We have also conducted further studies on other samples of Fe$_{1+x}$Te, where measurements were only recorded with the tip polarized along the crystal $b$ and $c$ axes. The data is shown in Table \ref{table3}. The intensity of the magnetic peak $\mathrm{I}(q_a)$ for field applied along the crystallographic $b$ and $c$ axes respectively and the corresponding canting angle ($\theta$) are shown. The average out-of-plane canting angle obtained from these values is $29.8\pm13.7\degree$. The error bar here contains contributions from variations in the magnetic properties of the STM tips, differences between samples and the alignment of the magnetic field plane with the crystallographic axis of the sample. This is the magnetic structure shown in Fig. \ref{fig:structure} $(b)$. 

\begin{figure}[t]
\includegraphics[width=90mm]{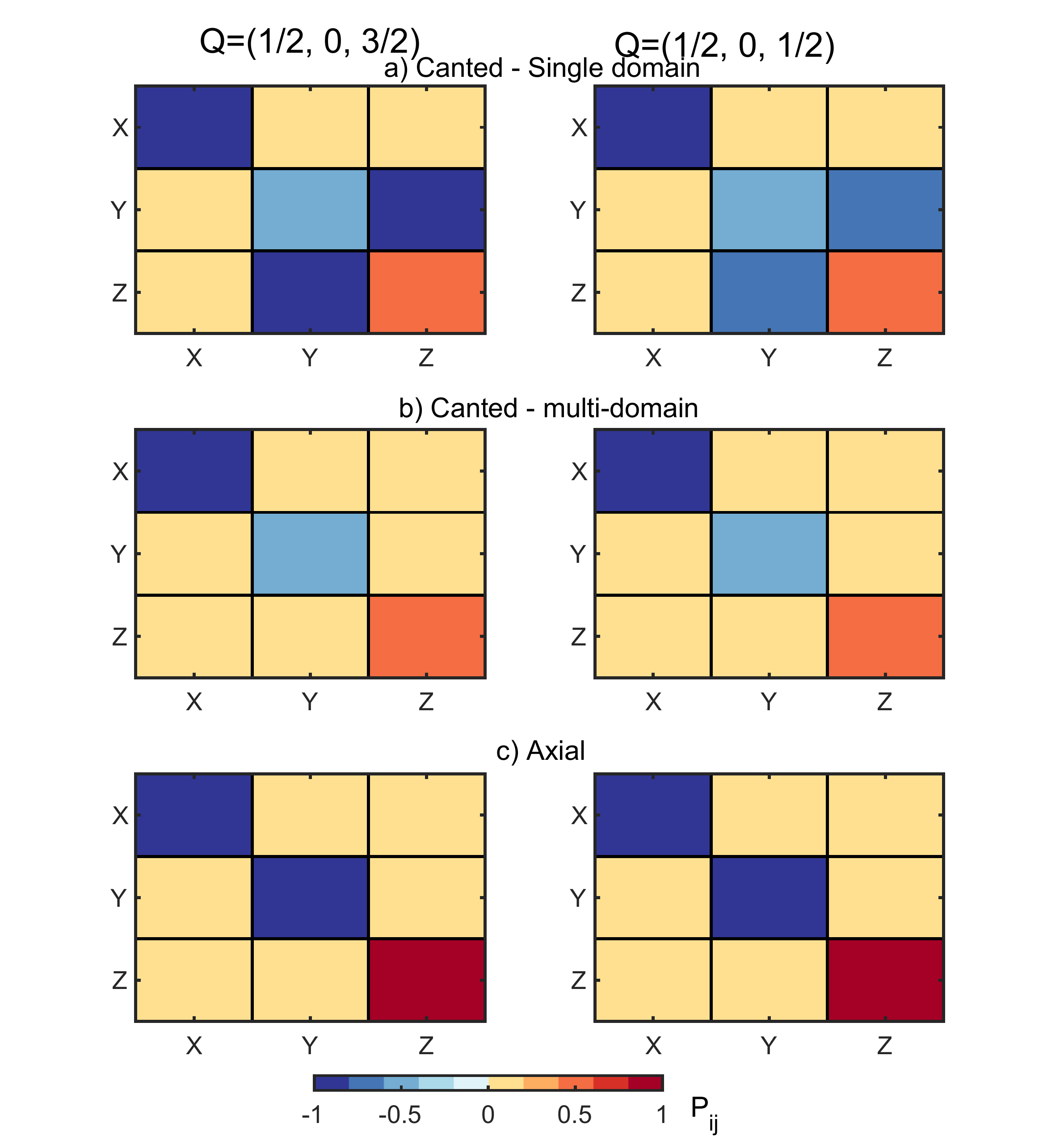}
\caption{A color schematic of the polarization matrix for the spin models under consideration for the $Q$-vectors $Q=(1/2,0,1/2)$ and $Q=(1/2,0,3/2)$: (a) for a single domain of the double-stripe order with canted spins with $\theta=28^\circ$, (b) for multiple domains of the same order as in (a) using the symmetry relations in table~\ref{table_pos} and (c) for the axial order with spins pointing along $b$. $X-$, $Y-$ and $Z-$ are the spin components along the three spatial directions, color encodes the polarization.}  \label{fig:matrix}
\end{figure}

\section{Neutron Scattering Experiments}

Having discussed the canted magnetic structure at the surface, we now apply neutron scattering to study the bulk magnetism.  Neutron scattering, unlike x-rays or photon based measurements, is a bulk measurement of materials owing to the interaction between neutrons and matter being mediated by nuclear interactions.  For example, for single crystalline Fe$_{1.09}$Te with a neutron wavelength of $\lambda=1.1\mathrm{\AA}$ the $1/e$ scattering length for the sum of absorption and incoherent cross sections is $\sim 2 \mathrm{cm}$.  The results presented in this section are therefore a measure of the bulk averaged response.  Neutrons carry a magnetic moment making them sensitive to the localized magnetic moments in materials.

\subsection{Experimental Details}

To investigate the polarization matrix of the magnetic order in Fe$_{1+x}$Te sensitive to the orientation of the local iron moments, we used the CRYOPAD (Cryogenic Polarization Analysis Device) developed at the ILL~\cite{Tasset99:267,Brown93:442}.  Unlike conventional polarization measurements which involve studying spin flip scattering along a particular crystallographic axis, CRYOPAD allows all components of the polarization matrix to be studied governed by the Blume-Maleev equations.~\cite{Blume63:130,Maleev63:4}  Single crystals of Fe$_{1.09}$Te were synthesized by the Bridgemann method~\cite{Rodriguez13:88,Rodriguez11:84}.  All measurements discussed here were done at a base temperature of $2\mathrm{K}$ using the IN20 spectrometer with the sample aligned such that Bragg peaks of the form (H 0 L) lay within the horizontal scattering plane.   The structural and magnetic transition in this material occurs at $\sim 60\mathrm{K}$ resulting in structural domains present at low temperatures.~\cite{Fobes14:112}  The possible symmetry operations resulting from these domains are displayed in Table \ref{table_pos} and discussed below.

\begin{table*}[ht]
\begin{tabular}{c|c|clcl}
\hline
$\vec{Q}$ & $P_{ij}^\mathrm{measured}$ & \ \ \ $P_{ij}^\mathrm{axial}$ & \ \ \ \ \ \ \ \ $ P_{ij}^\mathrm{canted}$ $(\theta=28^{\circ})$\\
\hline
$(\frac{1}{2},0,\frac{1}{2})$ & $ \left( \begin{array}{ccc} -0.896(3) & 0.064(4) & -0.002(4) \\ -0.086(3) & -0.879(3) & -0.018(4) \\ -0.049(4) & -0.032(4) & 0.882(3) \end{array}\right)$ & $\left( \begin{array}{ccc} -1 & 0 & 0 \\ 0 & -1 & 0 \\ 0 & 0 & 1  \end{array}\right)$ & $\left( \begin{array}{ccc} -1 & 0 & 0 \\ 0 & -0.768 & 0 \\ 0 & 0 & 0.768  \end{array}\right)$  \\
\hline
$(\frac{3}{2},0,\frac{3}{2})$  & $ \left( \begin{array}{ccc} -0.889(9) & 0.048(10) & -0.025(11) \\ -0.068(10) & -0.887(9) & -0.010(15) \\ -0.058(11) & -0.004(11) & 0.874(9) \end{array}\right)$ & $\left( \begin{array}{ccc} -1 & 0 & 0 \\ 0 & -1 & 0 \\ 0 & 0 & 1  \end{array}\right)$ & $\left( \begin{array}{ccc} -1 & 0 & 0 \\ 0 & -0.768 & 0 \\ 0 & 0 & 0.768  \end{array}\right)$  \\
\hline
$(\frac{3}{2},0,\frac{1}{2})$  & $ \left( \begin{array}{ccc} -0.887(9) & 0.042(10) & -0.039(11) \\ -0.071(10) & -0.877(9) & -0.016(15) \\ -0.042(11) & -0.020(10) & 0.887(9) \end{array}\right)$ & $\left( \begin{array}{ccc} -1 & 0 & 0 \\ 0 & -1 & 0 \\ 0 & 0 & 1  \end{array}\right)$ & $\left( \begin{array}{ccc} -1 & 0 & 0 \\ 0 & -0.636 & 0 \\ 0 & 0 & 0.636  \end{array}\right)$\\
\hline
$(\frac{1}{2},0,\frac{3}{2})$ & $ \left( \begin{array}{ccc} -0.887(3) & 0.068(4) & -0.003(4) \\ -0.060(3) & -0.880(3) & -0.016(4) \\ -0.048(4) & -0.031(4) & 0.884(4) \end{array}\right)$ & $\left( \begin{array}{ccc} -1 & 0 & 0 \\ 0 & -1 & 0 \\ 0 & 0 & 1  \end{array}\right)$ & $\left( \begin{array}{ccc} -1 & 0 & 0 \\ 0 & -0.946 & 0 \\ 0 & 0 & 0.946  \end{array}\right)$ \\
\hline
$(\frac{1}{2},0,\frac{5}{2})$ & $ \left( \begin{array}{ccc} -0.894(7) & 0.059(8) & -0.032(8) \\ -0.055(8) & -0.883(7) & -0.013(8) \\ -0.068(8) & -0.022(8) & 0.882(7) \end{array}\right)$ & $\left( \begin{array}{ccc} -1 & 0 & 0 \\ 0 & -1 & 0 \\ 0 & 0 & 1  \end{array}\right)$ & $\left( \begin{array}{ccc} -1 & 0 & 0 \\ 0 & -0.979 & 0 \\ 0 & 0 & 0.979  \end{array}\right)$\\
\hline
$(\frac{5}{2},0,\frac{1}{2})$ & $ \left( \begin{array}{ccc} -0.846(15) & 0.029(17) & -0.007(18) \\ -0.10(2) & -0.884(15) & -0.003(18) \\ -0.050(18) & -0.043(18) & 0.859(16) \end{array}\right)$ & $\left( \begin{array}{ccc} -1 & 0 & 0 \\ 0 & -1 & 0 \\ 0 & 0 & 1  \end{array}\right)$ & $\left( \begin{array}{ccc} -1 & 0 & 0 \\ 0 & -0.618 & 0 \\ 0 & 0 & 0.618  \end{array}\right)$\\
\hline
\end{tabular}
\caption{A list of the experimentally measured polarization matrices at $T= 2\mathrm{K}$ measured on IN20.  The calculated matrix elements~\cite{Qureshi19:52}, assuming 100 \% beam polarization, are shown for the axial spin structure and the canted magnetic structure with $\theta=28^\circ$ for comparison.}
\label{table}
\end{table*}

\begin{table}[ht]
\caption{Low temperature structural domains considered here for the magnetic structural analysis.}
\centering
\begin{tabular} {c c c c}
\hline
 1 & x & y &  z \\
 2 & x & y &  -z \\
 3 & -x & -y &  z \\
 4 & -x & -y &  -z \\
\hline
\label{table_pos}
\end{tabular}
\end{table}

Spherical neutron polarimetry is sensitive to the direction of the ordered magnetic moment, spin chirality, and coupling between nuclear and magnetic cross sections~\cite{Blume63:130}.  In the case of structural domains that exist at low temperature (which average out the off-diagonal elements) from the structural transition (Table \ref{table_pos}) and in the absence of spin chirality and coupling to a nuclear cross section, the polarization matrix measured with spherical neutron polarimetry becomes diagonal and takes the following form,

\begin{equation}
P_{ij} = 
\begin{pmatrix}
-1 & 0 &  0 \\
0 & {{|M_{\perp,y}|^{2}-|M_{\perp,z}|^{2}}\over|\vec{M}_{\perp}|^2} & 0 \\
0 & 0 &  -{{|M_{\perp,y}|^{2}-|M_{\perp,z}|^{2}}\over|\vec{M}_{\perp}|^2} 
\end{pmatrix} \nonumber
\end{equation}

\noindent where $\vec{M}_{\perp}\equiv \vec{Q} \times \vec{M} \times \vec{Q}$.   Here $\vec{Q} \equiv \vec{k}_{i}-\vec{k}_{f}$ is the momentum transfer and $\vec{M}$ is the magnetic moment direction. The matrix element $P_{xx} \equiv P_{11}$ is strictly $=-1$ and deviations from this are a measure of the neutron beam polarization.  It is important to note that the polarization matrix does not provide information on the magnitude of $\vec{M}$, but only the direction.

In this experiment, the polarization matrix was measured at six magnetic Bragg peaks at $T=2\mathrm{K}$.  The full experimental polarization matrices $P_{ij}^\mathrm{measured}$ for these Bragg peaks are shown in Table~\ref{table}.  The calculated matrices $P_{ij}^\mathrm{axial}$ and $P_{ij}^\mathrm{canted}$ shown in the table are discussed below in the context of our comparison with tunnelling and previous neutron results.

\subsection{Neutron scattering results}

Figure \ref{fig:structure} illustrates the two magnetic structures that we will compare the polarized neutron scattering results to in this section.  The reported structure based on neutron diffraction on single crystals and also powders suggests that the structure is collinear with the moments aligned along the crystallographic $b$ axis (Fig.~\ref{fig:structure} (a)).  The structure is often referred to as a ``double-stripe" magnetic structure.  This is contrasted to a recent magnetic structure reported using scanning tunnelling microscopy (Fig.~\ref{fig:structure} (b)).  The magnetic structure obtained from STM has the magnetic moments collinear but canted along the crystallographic $c$ axis by an angle of $\theta=29.8 \pm 13.7^{\circ}$.  For the purposes of this section we refer to the neutron scattering structure which is aligned along the $b$ axis as ``axial" and the structure reported by spin polarized tunnelling microscopy as ``canted".  We now discuss the application of neutron spherical polarimetry to revisit the bulk magnetic structure of collinear Fe$_{1+x}$Te.

\begin{figure}[t]
\includegraphics[width=90mm]{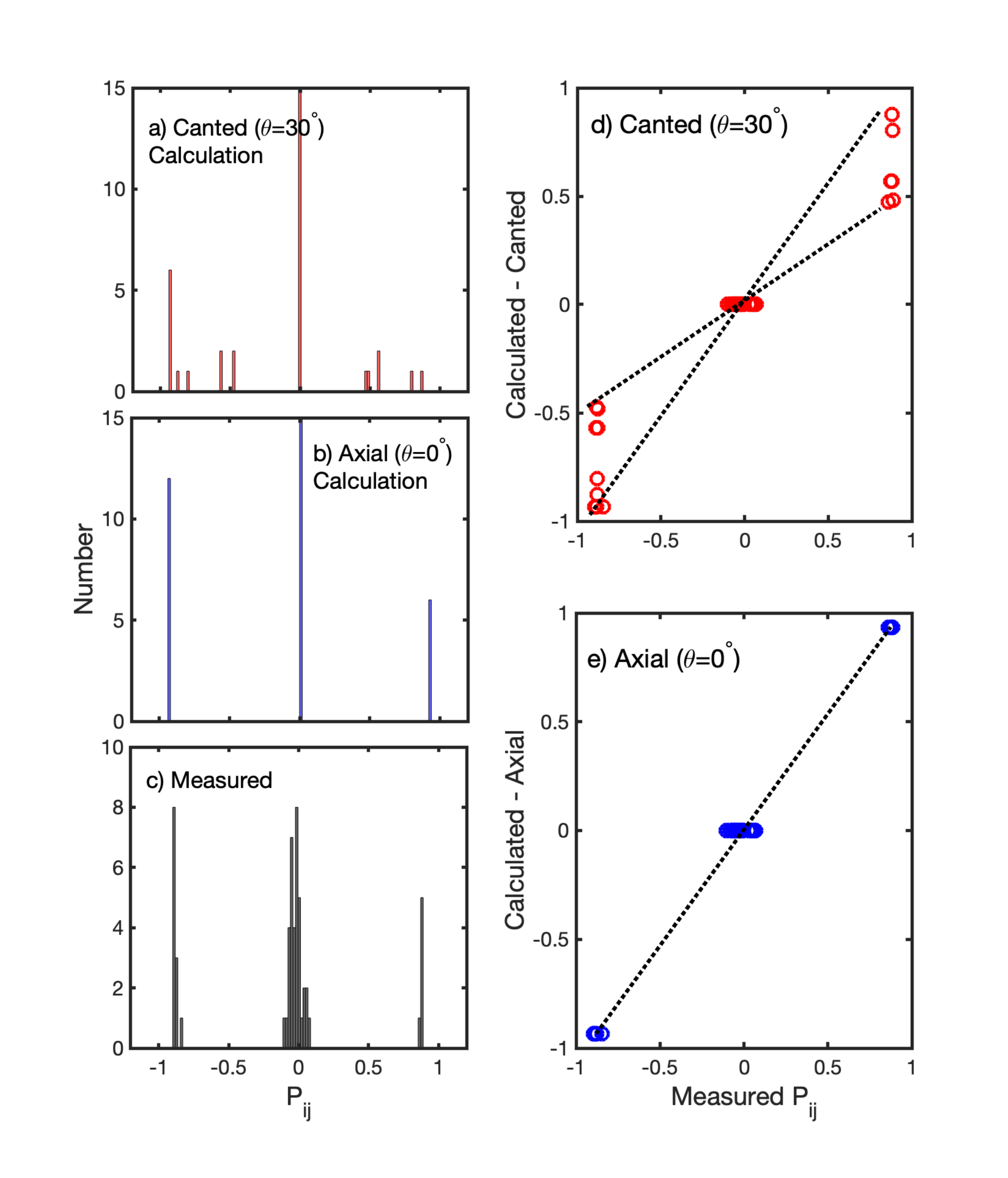}
\caption{A summary of the spherical polarimetry data from IN20 measured for Fe$_{1.09}$Te and compared against calculations. (a,b) histograms of the polarization matrix elements $P_{ij}$ for calculations based on the canted and axial magnetic structures respectively.  (c) the same histogram for the measured matrix elements.  (d,e) plots of the calculated polarization matrix elements as a function of measured values for both the canted and axial  magnetic structures.} \label{fig:data}
\end{figure}

\begin{figure}[t]
\includegraphics[width=90mm]{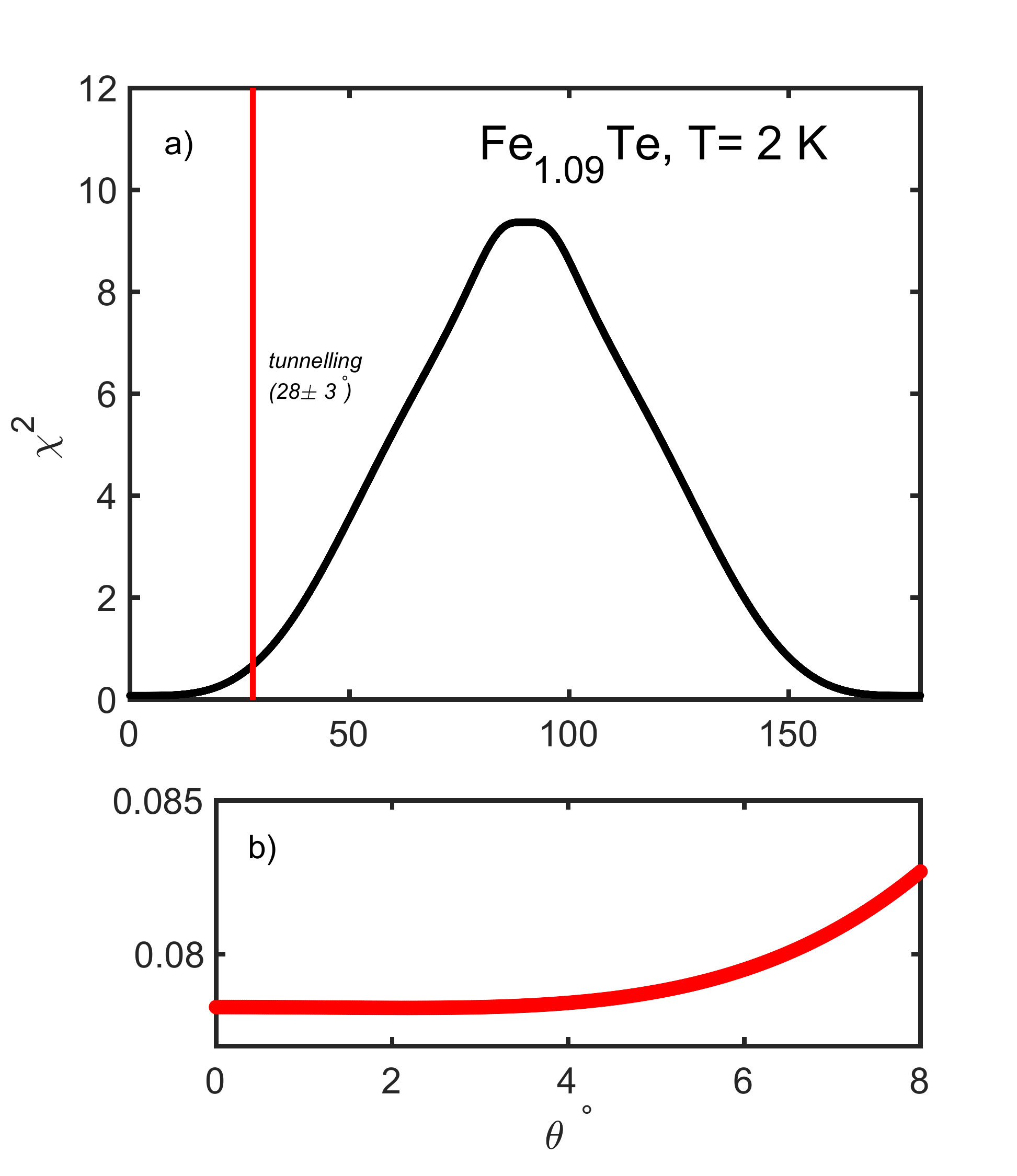}
\caption{A parameterization of the goodness of fit ($\chi^{2}$) to the data as a function of canting angle $\theta$ as defined in the text. A neutron beam polarization of $0.88$ was taken for the analysis. (a) $\chi^{2}$ over the full range of canted angles from $0-180^{\circ}$. (b) $\chi^2$ in a narrow range of angles from $0-5^\circ$.} \label{fig:chisq}
\end{figure}

While the application of unpolarized neutron powder diffraction and also uniaxial polarized neutrons maybe arguably ambiguous in determining canting of the magnetic moments owing to the number of accessible peaks and statistics for low interstitial iron concentrations, spherical neutron polarimetry is very sensitive to this canting.  We illustrate this in Fig. \ref{fig:matrix} which displays a color representation of the calculated polarization matrices at the magnetic momentum positions $\vec{Q}$=($\frac{1}{2}$, 0, $\frac{3}{2}$) and ($\frac{1}{2}$, 0, $\frac{1}{2}$).  Three different models are presented.  Panel $(a)$ displays a calculation based on the canted model proposed by tunnelling measurements for a $\textit{single}$ structural and magnetic domain crystal.  This calculation shows a non zero off-diagonal values for the matrix elements for the $P_{yz}$ and $P_{zy}$ positions.  However, Fe$_{1+x}$Te undergoes a structural distortion from a tetragonal to a monoclinic unit cell that is coincident with magnetic ordering.  The four domains are related by symmetry as displayed in Table~\ref{table_pos}.  The corresponding matrix including the effects of domains is diagonal and is illustrated in Fig.~\ref{fig:matrix}(b).  The magnitudes of the matrix elements $|P_{yy}| \neq |P_{zz}|$.  This contrasts with the case where the magnetic moments point within the $ab$ plane as termed ``axial" in this paper and schematically shown in Fig. \ref{fig:matrix}(c) where $|P_{yy}| = |P_{zz}|$.

Figure \ref{fig:data} illustrates a comparison of our results to the predicted matrix from both magnetic structures.  Figs. \ref{fig:data}(a, b) illustrate histograms of the calculated polarization matrix elements for the both the canted (tunnelling, Fig. \ref{fig:data}(a) and axial (neutron, Fig. \ref{fig:data}(b) magnetic structures.  The canted magnetic structure results in polarization matrix elements for a range of values ranging from $-1 \rightarrow 1$.  The largest number of matrix elements appear at $0$ resulting from the averaging over domains meaning that all off-diagonal matrix elements are calculated to be $0$ (Fig.~\ref{fig:matrix}).  The axial magnetic structure, in contrast only displays three matrix elements ($[-1, 0 ,1]$) distinguishing it from the canted magnetic structure.  

A histogram of the experimentally measured polarization matrix elements is plotted in Figs. \ref{fig:data}(c) and shows a distribution of measured elements centered around three values, in qualitative agreement with the axial (neutron) magnetic structure.  We note that there is a large distribution around $P_{ij}=0$ and is further shown in the table displayed above (Table \ref{table} in the experimental section).  The origin of this error results from the incomplete polarization of the beam and also due to small misalignments ($\sim 1-2^{\circ}$, see the appendix of Ref. \onlinecite{Giles_Donovan20:102} for an analysis of the errors) of the sample with respect to the beam polarization.   Based on the comparison between the Figs. \ref{fig:data} (a-c), the neutron data is consistent with the axial magnetic structure rather than the prediction of a broad spread of matrix elements which would result from a canted magnetic structure.  This is further illustrated in Figs. \ref{fig:data} (d,e) which shows the calculated matrix elements as a function of the measured matrix elements.  The spread of the data from a single straight line is a measure of the ``goodness of fit".  The canted magnetic structure in panel (c) clearly provides a much poorer description of the data over the axial one displayed in panel (d).

Figure \ref{fig:chisq} shows a plot of $\chi^{2}$ (a measure of the goodness of fit) as a function of canting angle $\theta$ quantifying the sensitivity of our measurement using spherical polarimetry and also establishing a measure of the errorbar in our experiment. For this figure, we have defined $\chi^{2}$ in terms of the measured and calculated polarization matrix elements $P_{i}$ by

\begin{equation}
\chi^{2} \equiv \sum_{i,j}\sum_{\alpha} (P^{\mathrm{measured},\alpha}_{ij}-P^{\mathrm{calculated},\alpha}_{ij})^{2},
\end{equation}

\noindent where the summation index $\alpha$ is taken over all Bragg peak peaks and the index $ij$ are the matrix elements probed in this experiment.  The plot of $\chi^{2}$ as a function of canting angle for the 6 Bragg peaks (Table \ref{table}) studied on IN20 shows a broad minimum near $\theta\sim 0^\circ$ and a distinct maximum with $\theta =90^{\circ}$ when the moments are pointing along the crystallographic $c$ axis.  The vertical red line in Fig. \ref{fig:chisq}(a) is the canted $\theta=28 \pm 3^{\circ}$ proposed by tunnelling measurements.  The $\chi^{2} (\theta)$ curve clearly shows that our spherical neutron polarimetry data is inconsistent with a canted structure with a broad minimum observed near $\theta=0^{\circ}$ which is the axial structure found previously in powders and single crystal unpolarized neutron measurements.  The nature of the broad minimum in the $\chi^{2}$ surface indicates an underlying errorbar in the magnetic structure measured here of $\sim$ $\pm 5^{\circ}$.  The neutron scattering data shows that the magnetic structure in iron deficient Fe$_{1.09}$Te is inconsistent with the magnitude of the canted magnetic structure reported for the single layer limit in tunnelling measurements.

\section{Discussion}
The comparison of the spherical neutron polarimetry with spin-polarized STM shows clearly that for Fe$_{1+x}$Te, a surface magnetic reconstruction forms where the spins on the iron site tilt out of the $ab$-plane. In the following we will discuss possible mechanisms leading to this reconstruction.

\subsection{Surface relaxation --  Density functional theory (DFT) calculations}

\begin{figure}[t]
\includegraphics[width=90mm]{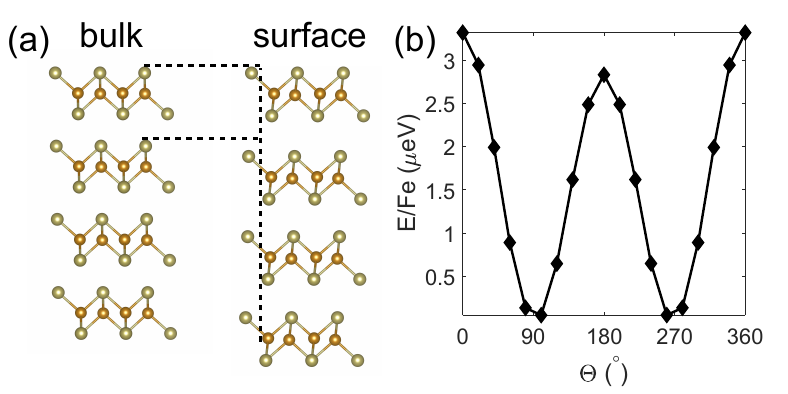}
\caption{(a) Structural model of the surface layer relaxed in DFT calculations. The relaxation of the top surface layer can be seen. (b) Calculation of the energy per Fe atom due to the dipole interaction as a function of out of plane canting angle $\theta$ of a bi-layer of $\mathrm{FeTe}$ as described in the text. The maximum energy per Fe atom due to the dipole interaction is found from this calculation to be $3.3 \mathrm{\mu eV}$. The energy minimum is found at $\theta=\pm90^\circ$, i.e. out of the surface plane.}  
\label{F2}
\end{figure}

To attempt to explain the difference between the measured magnetic structure of the surface and the bulk we have performed DFT calculations on an FeTe slab where the surface was allowed to relax from the lattice positions in the bulk. First-principles calculations were performed using the Quantum \textsc{Espresso}~\cite{Giannozzi_2009} code. We employed optimized norm-conserving Vanderbilt  pseudopotentials~\cite{hamann_optimized_2013} with the Perdew-Burke-Ernzerhof exchange-correlation functional in the generalized gradient approximation~\cite{perdew_generalized_1996}.  For the calculation, four layers of FeTe, without excess iron atoms ($x=0$), with a vacuum region of $13\mathrm{\AA}$ in the $z$ direction, bicollinear magnetic order along $x$, and ferromagnetic order along $z$ were considered. The $z$-length of the unit cell was kept fixed during variable cell relaxation runs and spin-polarization taken into account.  We chose a kinetic energy cutoff for the plane waves of $80\mathrm{Ry}$, a Methfessel-Paxton~\cite{Methfessel_1989} smearing of $0.02\mathrm{Ry}$, and a $8 \times 16 \times 1$ Monkhorst-Pack~\cite{Monkhorst_1976} \textbf{k}-mesh.  Details of the crystal structure and atomic positions were taken from experiment~\cite{Li09:79} and then geometrically relaxed. The surface layer was found to relax away from the bulk layer to such an extent that the $c$ axis parameter of the surface layer changes from $6.2\mathrm{\AA}$ to $6.8\mathrm{\AA}$. The lateral position of the surface layer also change from that of the bulk with the Fe atoms of the surface layer displaced by up to $10\mathrm{pm}$ from the unrelaxed position. Furthermore slight changes in the length of the $\mathrm{Fe}-\mathrm{Te}$ bonds of up to $3\mathrm{pm}$ were observed. This reconstructed structure is illustrated schematically in Fig.~\ref{F2}(a).

\subsection{Magnetic dipole interactions}

As a potential explanation for the canting of the spins in the surface layer, we have considered the magnetic dipole interaction between the Fe atoms. We have constructed a numerical model of the surface of Fe$_{1+x}$Te to calculate the preferred spin orientation of the surface Fe atoms given dipolar interactions with the layer below. The model consists of a bulk layer of $101\times101$ Fe atoms with a surface layer of $41\times41$ Fe atoms. The spins of each Fe atom were fixed into the bicollinear AFM order with their in-plane component fixed to point along the crystallographic $b$ axis. The magnetic moment of each Fe atom was taken to be $2\mu_\mathrm B$ \cite{Rodriguez11:84}. The energy of the interaction of the dipoles in the system is then determined by numerically summing over all the spins in the system. The equation for this process is:

\begin{equation}   E=\sum_{i\neq j}\frac{\mu_0}{4\pi|\bm{r}_{ij}|^2}\left[\bm{\mu}_i\cdot\bm{\mu}_j – \frac{3}{|\bm{r}_{ij}|^2}(\bm{\mu}_{i}\cdot\bm{r}_{ij})(\bm{\mu}_{j}\cdot\bm{r}_{ij})\right]
\label{dipole}
\end{equation}

\noindent where the indices $i$ and $j$ indicate the different $\mathrm{Fe}$ positions in both layers. By varying the canting angle $\theta$ of the surface spins we determined how the energy of the dipole interaction varies as a function of $\theta$, this is shown in Fig.~\ref{F2}(b). We determine from this analysis that the dipole interaction in Fe$_{1+x}$Te would favor aligning the Fe spins along the crystallographic $c$ axis, a result that is in good agreement with the effects of magnetic dipole interactions in bulk crystals \cite{Johnson2016}. Indeed if one were to consider only dipole interactions in addition to the AFM ordering in the bulk then the magnetic moments of the Fe would align with the sample $c$ axis. The fact that this is not observed in neutron scattering measurements leads us to the conclusion that a substantial magnetic anisotropy from the crystalline electric field keeps the Fe spins pointing in the $ab$ plane. We note that the tendency of the magnetic dipole interaction to favour out-of-plane order is stronger for the surface layer than it is in the bulk.  The energy scale associated with the dipolar interaction per Fe spin is $\sim \mu$eV. 


\subsection{Magnetocrystalline anisotropy}

The energy scale of the dipolar interaction is extremely small in comparison to the measured $\sim$ 5 meV anisotropy gap found with neutron inelastic scattering\cite{Stock11:84}.  Both, the direction and magnitude of magnetic dipole interactions suggest that these are not sufficiently strong to explain the out-of-plane tilting of the magnetization in the surface layer. This leaves the magnetocrystalline anisotropy resulting from crystalline electric field effects~\cite{Turner09:80,Yosida:book} as a possible origin for the magnetic surface reconstruction. In bulk Fe$_{1+x}$Te, the magnetic anisotropy results in an in-plane orientation of the spins\cite{Enayat14:345}. At the surface, the broken symmetry resulting from the loss of a mirror plane and structural relaxation of the surface layer discussed above imply that the magnetic anisotropy can differ significantly.

\section{Conclusions}

The spherical neutron polarimetry results show a distinct difference between the bulk magnetic structure in Fe$_{1.09}$Te measured with neutron scattering and the canted magnetic structure reported with tunnelling measurements in the single layer limit.  This illustrates a difference between bulk and surface magnetism in this Van der Waals magnet.  It should be noted that the cases of tunnelling from a surface and neutron scattering from the bulk are not studying the exact same situation.  In the bulk neutron response, each magnetic Fe$_{1+x}$Te layer effectively represents a mirror plane, this is not the case of a hard surface as is the situation in tunnelling.  Therefore, from a symmetry perspective, there is no constraint forcing both situations to be identical.  

The magnetic moments in Fe$_{1.09}$Te interact through either effects of bonding (including possible itinerant interactions such as RKKY exchange) or dipolar interactions.  For interactions within the $ab$ plane of Fe$_{1.09}$Te these should be dominated by the effects of bonding which result in strong dispersion of the magnetic excitations along these directions~\cite{Stock90:14}.  The situation along the $c$-axis is less clear as the FeTe layers are only weakly bonded through Van der Waals forces.  However, dipolar forces which decay $\sim {1\over r^{3}}$ are still present in the magnetic Hamiltonian and these could be strongly influential to the magnetic correlations along the crystallographic $c$-axis.  Magnetic neutron inelastic scattering have indeed found the weak $c$-axis correlations~\cite{Xu17:96,Stock11:84} occuring without the presence of strong bonding and only Van der Waals forces.

Another effect not directly tied to the crystalline electric field effects discussed above that may be the origin of the difference between surface (tunnelling) and bulk (neutron) responses is interstitial iron.  Previous neutron scattering results have shown a strong connection between the magnetic correlations and the interstitial iron concentration $x$.~\cite{Thampy14:90,Stock11:84}  With increasing interstitial iron concentration, the crystallographic $c$-axis decreases which could in turn increase the importance of the dipolar terms in the magnetic Hamiltonian.  The interstitial sites may also be magnetic and this could influence the structure in the FeTe plane.

We note that differences in the magnetic structure and periodicity between tunnelling and bulk neutrons scattering have been reported before.  Comparative measurements done in superconducting La$_{2-x}$Sr$_{x}$CuO$_{4}$~\cite{Christensen04:93} with tunnelling and neutron scattering have observed different wavevectors, however a similar response in the dynamics.  The role of dipolar and crystalline electric field terms in the magnetic Hamiltonian may be an issue that needs to be considered in all magnetic layered and two dimensional structures.

While further calculations will be required to ultimately understand the difference in magnetic structures observed on the surface and the bulk, our study illustrates the sensitivity and difference between the magnetism in the Fe$_{1.09}$Te Van der Waals magnet between the bulk and the surface.  This has been established through a comparison between spherical polarimetry to determine that the bulk magnetic structure of iron deficient Fe$_{1.09}$Te has $\theta$=0 $\pm$ 5$^{\circ}$, while using spin-polarized STM to characterize the surface magnetic order.  We suggest the difference between magnetic structures found between scanning tunnelling microscopy and neutron scattering originates from the relaxation of the surface layer and the corresponding changes in magnetocrystalline anisotropy.

We acknowledge financial support from the EPSRC (EP/R031924/1 and EP/R032130/1) and NIST Center for Neutron Research. C.H. acknowledges support by the Austrian Science Fund (FWF) Project No. P32144-N36 and the VSC4 of the Vienna University of Technology.

\end{document}